\newcount\Comments  
\documentclass{myarticle_EC}
\usepackage{bm}  
\renewcommand{\vec}[1]{\ensuremath{\bm{#1}}} 
\newcommand{\SP}{\mathsf{\# P}}
\renewcommand{\P}{\mathsf{P}}
\newcommand{\NP}{\mathsf{NP}}

\title{Complexity of Combinatorial Market Makers\titlenote{Part of this work was done
while L. Fortnow, N. Lambert and J. Wortman
were at Yahoo! Research, New York.}}
\numberofauthors{5}
\author{
\alignauthor
Yiling Chen\\
\affaddr{Yahoo!\ Research}\\
\affaddr{111 W. 40th St., 17th Floor}\\
\affaddr{New York, NY 10018}\\
\alignauthor
Lance Fortnow\\
\affaddr{EECS Department}
\affaddr{Northwestern University}\\
\affaddr{2133 Sheridan Road}\\
\affaddr{Evanston, IL 60208}\\
\alignauthor Nicolas Lambert\\
\affaddr{Department of Computer Science}\\
\affaddr{Stanford University}\\
\affaddr{Stanford, CA 94305}\\
\and \alignauthor
David M.\ Pennock\\
\affaddr{Yahoo!\ Research}\\
\affaddr{111 W. 40th St., 17th Floor}\\
\affaddr{New York, NY 10018}\\
\alignauthor
Jennifer Wortman\\
\affaddr{Department of Computer and Information Science}\\
\affaddr{University of Pennsylvania}\\
\affaddr{Philadelphia, PA 19104}\\
}

\begin{document}

%

\maketitle

\begin{abstract}
We analyze the computational complexity of market maker pricing algorithms for combinatorial prediction markets. We focus on Hanson's popular logarithmic market scoring rule market maker (LMSR). Our goal is to implicitly maintain correct LMSR prices across an exponentially large outcome space. We examine both permutation combinatorics, where outcomes are permutations of objects, and Boolean combinatorics, where outcomes are combinations of binary events. We look at three restrictive languages that limit what traders can bet on. Even with severely limited languages, we find that LMSR pricing is $\SP$-hard, even when the same language admits polynomial-time matching without the market maker. We then propose an approximation technique for pricing permutation markets based on a recent algorithm for online permutation learning.  The connections we draw between LMSR pricing and the vast literature on online learning with expert advice may be of independent interest.
\end{abstract}



\section{Introduction}
One way to elicit information is to ask people to bet on it. A \emph{prediction market} is a common forum where people bet with each other or with a market maker~\cite{Forsythe92,Forsythe99,Thaler88,Nelson89,Oliven04}. A typical binary prediction market allows bets along one dimension, for example either for or against Hillary Clinton to win the 2008 US Presidential election. Thousands of such one- or small-dimensional markets exist today, each operating independently. For example, at the racetrack, betting on a horse to win does not directly impact the odds for that horse to finish among the top two, as logically it should, because the two bet types are handled separately.

A \emph{combinatorial prediction market} is a central clearinghouse for handling logically-related bets defined on a combinatorial space. For example, the outcome space might be all $n!$ possible permutations of $n$ horses in a horse race, while bets are properties of permutations such as ``horse A finishes 3rd'' or ``horse A beats horse B.'' Alternately, the outcome space might be all $2^{50}$ possible
state-by-state results for the Democratic candidate in the 2008 US
Presidential election, while bets are Boolean statements such as
``Democrat wins in Ohio and Florida but not in Texas.''

Low liquidity marginalizes the value of prediction markets, and
combinatorics only exacerbates the problem by dividing traders'
attention among an exponential number of outcomes. A combinatorial
matching market---the combinatorial generalization of a standard
double auction---may simply fail to find any
trades~\cite{Fortnow04,Chen07,Chen07-3}.

In contrast, an \emph{automated market maker} is always willing to
trade on \emph{every} bet at some price. A combinatorial market maker
implicitly or explicitly maintains prices across all (exponentially many) outcomes, thus allowing any trader at any time to place any bet, if transacted at the market maker's quoted price.

Hanson's~\cite{Hanson03,Hanson07} logarithmic market scoring rule
market maker (LMSR) is becoming the de facto standard market maker for
prediction markets. LMSR has a number of desirable properties,
including bounded loss that grows logarithmically in the number of
outcomes, infinite liquidity, and modularity that respects some
independence relationships. LMSR is used by a number of companies,
including inklingmarkets.com, Microsoft, thewsx.com, and yoonew.com,
and is the subject of a number of research
studies~\cite{Chen07-2, Hanson08, Chen07-4}.

In this paper, we analyze the computational complexity of LMSR in
several combinatorial betting scenarios. We examine both permutation
combinatorics and Boolean combinatorics. We show that both computing instantaneous prices and computing payments of transactions are $\SP$-hard in all cases we examine, even when we restrict
participants to very simplistic and limited types of bets. For
example, in the horse race analogy, if participants can place bets
only of the form ``horse A finishes in position N'',
then pricing these bets properly according to LMSR is $\SP$-hard, even
though matching up bets of the exact same form (with no market maker)
is polynomial~\cite{Chen07}.

On a more positive note, we examine an approximation algorithm for LMSR pricing in permutation markets that makes use of powerful techniques from the literature on online learning with expert advice~\cite{Cesa06,Littlestone94,Freund97}.  We briefly review this online learning setting, and examine the parallels that exist between LMSR pricing and standard algorithms for learning with expert advice.  We then show how a recent algorithm for permutation learning~\cite{Helmbold07} can be transformed into an approximation algorithm for pricing in permutation markets in which the market maker is guaranteed to have bounded loss.

\section{Related work}

Fortnow et al.~\cite{Fortnow04} study the computational complexity of
finding acceptable trades among a set of bids in a Boolean
combinatorial market. In their setting, the center is an
\emph{auctioneer} who takes no risk, only matching together willing
traders. They study a call market setting in which bids are collected
together and processed once en masse.  They show that the auctioneer
matching problem is co-NP-complete when orders are divisible and
$\Sigma_2^p$-complete when orders are indivisible, but identify a
tractable special case in which participants are restricted to bet on
disjunctions of positive events or single negative events.

Chen et al.~\cite{Chen07} analyze the the auctioneer matching problem
for betting on permutations, examining two bidding languages. \emph{Subset bets} are bets of the form ``candidate $i$ finishes in positions $x$, $y$, or $z$'' or ``candidate $i$, $j$, or $k$ finishes in position $x$.'' \emph{Pair bets} are of the form ``candidate $i$ beats candidate $j$.'' They give a polynomial-time algorithm for matching divisible subset bets, but show that matching pair bets is NP-hard.

Hanson highlights the use of LMSR for Boolean combinatorial markets,
noting that the subsidy required to run a combinatorial market on $2^n$
outcomes is no greater than that required to run $n$ independent
one-dimensional markets~\cite{Hanson03,Hanson07}.
Hanson discusses the computational difficulty of maintaining LMSR
prices on a combinatorial space, and proposes some solutions,
including running market makers on overlapping subsets of events,
allowing traders to synchronize the markets via arbitrage.

The work closest to our own is that of Chen, Goel, and Pennock~\cite{Chen08},
who study a special case of Boolean combinatorics in which participants bet on how far a team will advance in a single elimination tournament, for example a sports playoff like the NCAA college basketball tournament. They provide a polynomial-time algorithm for LMSR pricing in this setting based on a Bayesian network representation of prices. They also show that LMSR pricing is NP-hard for a very general bidding language. They suggest an approximation scheme based on Monte Carlo simulation or importance sampling.


We believe ours are the first non-trivial hardness results and worst-case bounded approximation scheme for LMSR pricing.

\section{Background}

\subsection{Logarithmic Market Scoring Rules}
Proposed by Hanson \cite{Hanson03,Hanson07}, a logarithmic market
scoring rule is an automated market maker mechanism that always
maintains a consistent probability distribution over an outcome
space $\Omega$ reflecting the market's estimate of the likelihood of each outcome.  A generic LMSR offers a security corresponding to each possible outcome $\omega$. The security associated to outcome $\omega$ pays off \$1 if the outcome $\omega$ happens, and \$0 otherwise. Let
$\vec{q}=(q_{\omega})_{\omega \in \Omega}$ indicate the number of
outstanding shares for all securities. The LMSR market maker starts
the market with some initial shares of securities, $\vec{q^0}$,
which may be $\vec{0}$. The market keeps track of the outstanding shares of securities $\vec{q}$ at all times, and maintains a cost function
\begin{equation}\label{E:lmsr-cost}
C(\vec{q}) = b \log \sum_{\omega \in \Omega} \e^{q_{\omega}/b},
\end{equation}
and an instantaneous price function for each security
\begin{equation}\label{E:lmsr-price}
p_{\omega}(\vec{q})= \frac{ \e^{q_{\omega}/b}}{\sum_{\tau \in \Omega}
\e^{q_{\tau}/b}},
\end{equation}
where $b$ is a positive parameter related to the depth of the
market. The cost function captures the total money wagered in the
market, and $C(\vec{q^0})$ reflects the market maker's maximum subsidy to the market. The instantaneous price function $p_{\omega}(\vec{q})$
gives the current cost of buying an infinitely small quantity of the
security for outcome $\omega$, and is the partial derivative of the
cost function, i.e. $p_{\omega}(\vec{q}) = \partial
C(\vec{q})/\partial q_{\omega}$. We use
$\vec{p}=(p_{\omega}(\vec{q}))_{\omega \in \Omega}$ to denote the
price vector. Traders buy and sell securities through the market
maker. If a trader wishes to change the number of outstanding shares
from $\vec{q}$ to $\vec{\tilde{q}}$, the cost of the transaction
that the trader pays is $C(\vec{\tilde{q}}) - C(\vec{q})$, which
equals the integral of the price functions following any path from $\vec{q}$ to
$\vec{\tilde{q}}$.

When the outcome space is large, it is often natural to offer only compound securities on sets of outcomes. A compound security $\set$ pays \$1 if one of the outcomes in the set $\set \subset \Omega$ occurs and \$0 otherwise.
Such a security is the combination of all securities $\omega \in \set$. Buying or selling $q$ shares of the compound security $\set$ is equivalent to buying or selling $q$ shares of each security $\omega \in \set$.  Let $\Theta$ denote the set of all allowable compound securities. Denote the outstanding shares of all compound securities as $Q = (q_{S})_{S\in\Theta}$. The cost function can be written as
\begin{eqnarray}\label{E:lmsr-cost2}
C(Q)&=&b \log \sum_{\omega \in \Omega}\e^{\sum_{\set\in\Theta:\omega\in\set}q_{S}/b}
\nonumber
\\
&=&b \log \sum_{\omega \in \Omega}\prod_{\set\in\Theta:\omega\in\set}\e^{q_{S}/b}~.
\end{eqnarray}
The instantaneous price of a compound security $\set$ is computed as the sum of the
instantaneous prices of the securities that compose the compound security $\set$,
\begin{eqnarray}
\price_{\set}(Q)
&=& \frac{\sum_{\state \in \set} \e^{\quant_{\state}/b}}
{\sum_{\tau \in \states} \e^{\quant_{\tau}/b}}
\nonumber
\\
&=& \frac{\sum_{\state \in \set} \e^{\sum_{\set'\in\Theta:\state\in \set'} \quant_{\set'}  /b}}
{\sum_{\tau\in \states} \e^{\sum_{\set'\in\Theta:\tau\in \set'} \quant_{\set'}  /b}}
\nonumber
\\
&=& \frac{\sum_{\state \in \set} \prod_{\set'\in\Theta:\state\in \set'} \e^{\quant_{\set'}  /b}}
{\sum_{\tau\in \states} \prod_{\set'\in\Theta:\tau\in \set'} \e^{\quant_{\set'}  /b}} ~.
\label{eq:lmsr}
\end{eqnarray}

Logarithmic market scoring rules are so named because they are based
on {\em logarithmic scoring rules}. A logarithmic scoring rule is a
set of reward functions
\[
 \{s_{\omega}(\vec{r})=a_{\omega} + b
\log(r_{\omega}): \omega \in \Omega\} ,
\]
where $\vec{r}=(r_{\omega})_{\omega \in \Omega}$ is a probability distribution over $\Omega$, and
$a_{\omega}$ is a free parameter. An agent who reports $\vec{r}$ is
rewarded $s_{\omega}(\vec{r})$ if outcome $\omega$ happens.
Logarithmic scoring rules are {\em proper} in the sense that when
facing them a risk-neutral agent will truthfully report his
subjective probability distribution to maximize his expected reward.
A LMSR market can be viewed as a sequential version of logarithmic
scoring rule, because by changing market prices from $\vec{p}$ to
$\vec{\tilde{p}}$ a trader's net profit is
$s_{\omega}(\vec{\tilde{p}}) - s_{\omega}(\vec{p})$ when outcome
$\omega$ happens. At any time, a trader in a LMSR market is essentially facing a logarithmic scoring rule.

LMSR markets have many desirable properties. They offer consistent pricing for
combinatorial events. As market maker mechanisms, they provide
infinite liquidity by allowing trades at any time. Although the market
maker subsidizes the market, he is guaranteed a worst-case loss no
greater than $C(\vec{q^0})$, which is $b\log n$ if $|\Omega|=n$ and the market starts with 0 share of every security. In addition, it is a dominant strategy
for a myopic risk-neutral trader to reveal his probability
distribution truthfully since he faces a proper scoring
rule. Even for forward-looking traders, truthful reporting is an
equilibrium strategy when traders' private information is
independent conditional on the true outcome~\cite{Chen07-2}.

\subsection{Complexity of Counting}

The well-known class $\NP$ contains questions that ask whether a search problem has a solution, such as whether a graph is 3-colorable. The class $\SP$ consists of functions that \emph{count} the number of solutions of $\NP$ search questions, such as the number of 3-colorings of a graph.

A function $g$ is $\SP$-hard if, for every function $f$ in $\SP$, it is possible to compute $f$ in polynomial time given an oracle for $g$. Clearly oracle access to such a function $g$ could additionally be used to solve any $\NP$ problem, but in fact one can solve much harder problems too. Toda~\cite{toda} showed that every language in the polynomial-time hierarchy can be solved efficiently with access to a
$\SP$-hard function.

To show a function $g$ is a $\SP$-hard function, it is sufficient to
show that a function $f$ reduces to $g$ where $f$ was previously
known to be $\SP$-hard. In this paper we use the following
$\SP$-hard functions to reduce from:
\begin{itemize}
\item {\bf Permanent}: The permanent of an $n$-by-$n$ matrix $A=(a_{i,j})$ is defined as
\begin{equation}\label{eq:perm}
\perman(A) = \sum_{\perm \in \Omega} \prod_{i=1}^n a_{i,\perm(i)} ~, \end{equation}
where $\Omega$ is the set of all permutations over $\{1, 2, ... , n\}$. Computing the permanent of a matrix $A$ containing $0$-$1$ entries is $\SP$-hard~\cite{Va}.
\item {\bf $\#$2-SAT}: Counting the number of satisfying assignments of a formula given in conjunctive normal form with each clause having two literals is $\SP$-hard~\cite{Valiant79}.
\item {\bf Counting Linear Extensions}: Counting the number of total orders that extend a partial order given by a directed graph is $\SP$-hard~\cite{Brightwell91}.
\end{itemize}
$\SP$-hardness is the best we can achieve since all the functions in
this paper can themselves be reduced to some other $\SP$ function.

\section{LMSR for Permutation Betting}

In this section we consider a particular type of market combinatorics in which the final outcome is a ranking
over $n$ competing candidates. Let the set of candidates be $\scN_n = \{1, \dots, n\}$, which is also used to represent the set of positions. In the setting, $\Omega$ is the set
of all permutations over $\scN_n$. An outcome $\sigma \in \Omega$ is
interpreted as the scenario in which each candidate $i$ ends up in position $\sigma(i)$.
Chen et al.~\cite{Chen07} propose two betting languages, \emph{subset betting} and \emph{pair betting}, for
this type of combinatorics and analyze the complexity of the auctioneer's order matching problem for each.
In what follows we address the complexity of operating an LMSR market for both betting
languages.

\subsection{Subset Betting}
\label{sec:subsetbetting}

As in Chen et al.~\cite{Chen07}, participants in a LMSR market for subset betting may trade two types of compound securities:  (1) a security of the form $\langle i | \Phi\rangle$ where $\Phi \subset \scN_n$ is a subset of positions; and (2) a security $\langle \Psi | j \rangle$ where $\Psi \subset\scN_n$ is a subset of candidates.  The security $\angles{i | \Phi}$ pays off \$1 if
candidate $i$ stands at a position that is an element of $\Phi$ and \$0 otherwise.  Similarly, the security $\angles{\Psi | j}$ pays off \$1 if any of the candidates in $\Psi$ finishes at position $j$ and \$0 otherwise. For example, in a horse race, participants can trade securities of the form  ``horse A will come in the second, fourth, or fifth place'', or ``either horse B or horse C will come in the third place''.

Note that owning one share of $\langle i | \Phi \rangle$ is equivalent to owning one share of  $\langle i | j \rangle$ for every $j \in \Phi$, and similarly owning one share of $\langle \Psi | j \rangle$ is
equivalent to owing one share of $\langle i | j \rangle$ for every
$i \in \Psi$. We restrict our attention to a simplified market
where securities traded are of the form $\langle i | j \rangle$. We show that even in this simplified market it is $\SP$-hard for the market maker to provide the instantaneous security prices, evaluate the cost function, or calculate payments for transactions, which implies that the running an LMSR market for the more general case of subset betting is also $\SP$-hard.

Traders can trade securities $\langle i | j \rangle$ for all $i \in \scN_n$ and $j \in \scN_n$ with the market maker.
Let $q_{i,j}$ be the total number of outstanding shares for security $\langle i | j \rangle$ in the market. Let  $Q=(q_{i,j})_{i \in \scN_n, j \in \scN_n}$ denote the outstanding shares for all securities. The market maker keeps track of $Q$ at all times. From Equation~\ref{eq:lmsr}, the instantaneous price of security $\langle i | j \rangle$ is
\begin{equation}\label{E:subsetprice}
\price_{i,j}(Q)
= \frac{\sum_{\perm \in \Omega: \perm(i) = j} \prod_{k=1}^n\e^{
\quant_{k,\perm(k)}  /b}} {\sum_{\tau \in \Omega}\prod_{k=1}^n\e^{
\quant_{k,\tau(k)}  /b}} ~,
\end{equation}
and from Equation~\ref{E:lmsr-cost2}, the cost function for subset betting is
\begin{equation}\label{E:subsetcost}
C(Q) = b \log \sum_{\perm \in \Omega}\prod_{k=1}^n\e^{
\quant_{k,\perm(k)}  /b}~.
\end{equation}

We will show that computing instantaneous prices, the cost function, and/or payments of transactions for a subset betting market is $\SP$-hard by a
reduction from the problem of computing the permanent of a (0,1)-matrix.
\begin{theorem}\label{T:subsetprice}
It is $\SP$-hard to compute instantaneous prices in a LMSR market for subset betting.  Additionally, it is $\SP$-hard to compute the value of the cost function.
\label{thm:subsethard}
\end{theorem}
\begin{proof}
We show that if we could compute the instantaneous prices or the value of the cost function for subset
betting for any quantities of shares purchased, then we could
compute the permanent of any $(0,1)$-matrix in polynomial time.

Let $n$ be the number of candidates, $A = (a_{i,j})$ be any
$n$-by-$n$ (0,1)-matrix, and $N = n!+1$. Note that
$\prod_{i=1}^n a_{i, \sigma(i)}$ is either 0 or 1. From Equation~\ref{eq:perm}, $\perman(A) \leq n!$ and hence $\perman(A) \mod N =
\perman(A)$. We show how to compute $\perman(A) \mod N$ from prices in subset betting markets in which $q_{i,j}$ shares of $\langle i | j \rangle$ have been purchased, where $q_{i,j}$ is
defined by
\begin{equation}\label{E:quan}
\quant_{i,j} =
\begin{cases}
b \ln N & \text{if $a_{i,j} = 0$,}
\\
b \ln (N+1) &\text{if $a_{i,j} = 1$}
\end{cases}
\end{equation}
for any $i \in \scN_n$ and any $j \in \scN_n$.

Let $B = (b_{i,j})$ be a $n$-by-$n$ matrix containing entries of the form
$b_{i,j} = \e^{\quant_{i,j}/b}$. Note that $b_{i,j} = N$ if $a_{i,j}
= 0$ and $b_{i,j} = N + 1$ if $a_{i,j} = 1$. Thus, $\perman(A)
\mod N = \perman(B) \mod N$.
Thus, from Equation~\ref{E:subsetprice}, the price for $\langle i | j \rangle$ in the market is
\begin{eqnarray*}
\price_{i,j} (Q) &=& \frac{\sum_{\perm \in \Omega: \perm(i) = j}  \prod_{k=1}^n
b_{k,\perm(k)}} {\sum_{\tau \in \Omega}  \prod_{k=1}^n
b_{k,\tau(k)}}
\\
&=& \frac{b_{i,j} \sum_{\perm \in \Omega: \perm(i) = j} \prod_{k\neq i}
b_{k,\perm(k)}} {\sum_{\tau  \in \Omega}  \prod_{k=1}^n
b_{k,\tau(k)}}
\\
&=& \frac{b_{i,j} \cdot \perman(M_{i,j})}{\perman(B)}
\end{eqnarray*}
where $M_{i,j}$ is the matrix obtained from $B$ by removing the
$i$th row and $j$th column.  Thus the ability to efficiently compute
prices gives us the ability to efficiently compute $\perman(M_{i,j})
/ \perman(B)$. 

It remains to show that we can use this ability to
compute $\perman(B)$.  We do so by telescoping a sequence of prices.  Let $B_i$ be the matrix $B$ with the first $i$ rows and columns removed. From above, we have $\perman(B_1) /\perman(B) = \price_{1,1}(Q)/b_{1,1}$.
Define $Q_m$ to be the $(n-m)$-by-$(n-m)$ matrix $(q_{i,j})_{i > m,j > m}$,
that is, the matrix of quantities of securities $(q_{i,j})$ with the
first $k$ rows and columns removed. In a market with only $n-m$ candidates, applying the same technique to the matrix $Q_m$, we can obtain $\perman(B_{m+1}) /
\perman(B_m)$ from market prices for $m = 1, ..., (n-2)$.
Thus by computing $n-1$ prices,
we can compute
\begin{eqnarray*}
\left(\frac{\perman(B_1)}{\perman(B)}\right)
\left(\frac{\perman(B_2)}{\perman(B_1)}\right)
&\cdots& \left(\frac{\perman(B_{n-1})}{\perman(B_{n-2})}\right)
\\
&=& \left(\frac{\perman(B_{n-1})}{\perman(B)}\right).
\end{eqnarray*}
Noting that $B_{n-1}$ only has one element, we thus can compute $\perman(B)$ from market prices.  Consequently, $\perman(B) \mod N$ gives $\perman(A)$.

Therefore, given a $n$-by-$n$ $(0,1)$-matrix $A$, we can compute the
permanent of $A$ in polynomial time using prices in $n-1$ subset
betting markets wherein an appropriate quantity of securities have
been purchased.

Additionally, note that
\[
C(Q) =b \log \sum_{\perm \in \Omega} \prod_{k=1}^n b_{k, \perm(k)} = b \log \perman(B)~.
\]
Thus if we can compute $C(Q)$, we can also compute $\perman(A)$.

As computing the
permanent of a $(0,1)$-matrix is $\SP$-hard, both computing market prices and computing the cost function in
a subset betting market are $\SP$-hard.
\end{proof}

\begin{corollary}\label{cor:subset}
Computing the payment of a transaction in a LMSR for subset betting is $\SP$-hard.
\end{corollary}
\begin{proof}
Suppose the market maker starts the market with 0 share of every security. Denote $Q^0$ as the initial quantities of all securities. If the market maker can compute $C(\tilde{Q}) - C(Q)$ for any quantities $\tilde{Q}$ and $Q$, it can compute $C(Q) - C(Q^0)$ for any $Q$. As $C(Q^0) = b\log n!$, the market maker is able to compute $C(Q)$. According to Theorem \ref{T:subsetprice}, computing the payment of a transaction is $\SP$-hard.
\end{proof}

\subsection{Pair Betting}
\label{sec:pairbetting}

In contrast to subset betting, where traders bet on absolute
positions for a candidate, pair betting allows traders to bet on the
relative position of a candidate with respect to another. More
specifically, traders buy and sell securities of the form $\langle i
> j \rangle$, where $i$ and $j$ are candidates. The security pays
off \$1 if candidate $i$ ranks higher than candidate $j$
(\ie $\sigma(i) < \sigma(j)$ where $\sigma$ is the final ranking of candidates) and \$0
otherwise. For example, traders may bet on events of the form
``horse A beats horse B'', or ``candidate C receives more votes than candidate D''.

As for subset betting, the current state of the market is determined by the total number of outstanding shares for all securities. Let $q_{i,j}$ denote the number of outstanding shares for $\langle i
> j \rangle$. Applying Equations~\ref{E:lmsr-cost2} and~\ref{eq:lmsr} to the present context, we find that the instantaneous price of
the security $\angles{i,j}$ is given by
\begin{equation}
\price_{i,j}(Q) = \frac{\sum_{\perm \in \Omega: \perm(i) < \perm(j)} \prod_{i',j'
: \perm(i') < \perm(j')} \e^{\quant_{i',j'} / b}} {\sum_{\tau \in \Omega}
\prod_{i',j' : \tau(i') < \tau(j')} \e^{\quant_{i',j'} / b}} ~,
\label{eqn:pairprice}
\end{equation}
and the cost function for pair betting is
\begin{equation}
C(Q) = b \log \sum_{\perm \in \Omega}
\prod_{i,j : \perm(i) < \perm(j)} \e^{\quant_{i,j} / b}~.
\end{equation}
We will show that computing prices, the value of the cost function, and/or payments of transactions for pair betting is $\SP$-hard via a reduction from the problem of computing the number of linear extensions to any
partial ordering.

\begin{theorem}\label{T:pairprice}
It is $\SP$-hard to compute instantaneous prices in a LMSR market for pair betting.  Additionally, it is $\SP$-hard to compute the value of the cost function.
\label{thm:pairhard}
\end{theorem}
\begin{proof}
Let $P$ be a partial order over $\{ 1, \dots, n \}$. We recall that a
linear (or total) order $T$ is a linear extension of $P$ if whenever $x \leq y$
in $P$ it also holds that $x \leq y$ in $T$. We denote by
$\linexts(P)$ the number of linear extensions of $P$.

Recall that $(i,j)$ is a covering pair of $P$ if $i \leq j$
in $P$ and there does not exist $\ell \neq i,j$ such that $i \leq \ell
\leq j$. Let $\{ (i_1,j_1)$, $(i_2, j_2)$, ... , $(i_k, j_k)\}$ be a
set of covering pairs of $P$. Note that covering pairs of a
partially ordered set with $n$ elements can be easily obtained in
polynomial time, and that their number is less than $n^2$.

We will show that we can design a sequence of trades that, given a
list of covering pairs for $P$, provide $\linexts(P)$ through a simple function of market prices.

We consider a pair betting market over $n$ candidates. We construct
a sequence of $k$ trading periods, and denote by $q^t_{i,j}$ and
$p^t_{i,j}$ respectively the outstanding quantity of security $\langle i
> j \rangle$
and its instantaneous price at the end of period $t$. At the beginning of the market, $q^0_{i,j} = 0$ for any $i$ and $j$. At each period t, $0<t\leq k$, $b \ln n!$ shares of security $\langle
i_t > j_t \rangle$ are purchased.

Let
\begin{equation*}
N_t(i,j) = \sum_{\perm \in \Omega: \perm(i) < \perm(j)} \prod_{i',j' :
\perm(i') < \perm(j')} \e^{\quant^t_{i',j'} / b}~,
\end{equation*}
and
\begin{equation*}
D_t = \sum_{\perm \in \Omega} \prod_{i',j' : \perm(i') < \perm(j')}
\e^{\quant^t_{i',j'} / b}~.
\end{equation*}
Note that according to Equation~\ref{eqn:pairprice}, $p^t_{i_t,j_t}
= N_t(i_t, j_t)/D_t$.

For the first period, as only the security $\langle i_1 > j_1 \rangle$ is purchased, we get
\begin{equation*}
D_1 = \sum_{\perm \in \Omega : \perm(i_1) < \perm(j_1)} n! + \sum_{\perm :
\perm(i_1) > \perm(j_1)} 1 = \frac{(n!)^2 + n!}{2} ~.
\end{equation*}

We now show that $D_k$ can be calculated inductively from $D_1$
using successive prices given by the market.
During period $t$, $b \ln n!$ shares of $\langle i_t > j_t \rangle$ are purchased. Note also that the securities purchased are different at each period, so that $q^s_{i_t,j_t} =0 $ if $s<t$ and $q^s_{i_t,j_t} = b \ln n!$ if $s \geq t$. We have
\begin{equation*}
N_t(i_t, j_t) = N_{t-1}(i_{t}, j_{t}) \e^{b \ln(n!) / b} = n!
N_{t-1}(i_{t}, j_{t})~.
\end{equation*}
Hence,
\begin{equation*}
\frac{p^{t}_{i_t, j_t}}{p^{t-1}_{i_t, j_t}}
=\frac{N_t(i_t, j_t) / D_t}{N_{t-1}(i_{t}, j_{t}) / D_{t-1}}
=\frac{n!D_{t-1}}{D_t} ~,
\end{equation*}
and therefore,
\begin{equation*}
D_k = (n!)^{k-1} \left(\prod_{\ell=2}^{k} \frac{p^{\ell - 1}_{i_\ell, j_\ell}}{p^\ell_{i_\ell, j_\ell}}\right)D_1~.
\end{equation*}
So $D_k$ can be computed in polynomial time in $n$ from the prices.

Alternately, since the cost function at the end of period $k$ can be written as $C(Q) = b\log D_k$, $D_k$ can also be computed efficiently from the cost function in period $k$.

We finally show that given $D_k$, we can compute $\linexts(P)$ in
polynomial time.
Note that at the end of the $k$ trading periods, the securities
purchased correspond to the covering pairs of $P$, such that
$\e^{q^k_{i,j}/b} = n!$ if $(i, j)$ is a covering pair of $P$ and $\e^{q^k_{i,j}/b} = 1$ otherwise.
Consequently, for a permutation $\perm$ that satisfies the partial
order $P$, meaning that $\perm(i) \leq \perm(j)$ whenever $i \leq j$
in P, we have
\begin{equation*}
\prod_{i',j' : \perm(i') < \perm(j')} \e^{\quant^k_{i',j'} / b} = (n!)^k ~.
\end{equation*}
On the other hand, if a permutation $\perm$ does not satisfy $P$, it
does not satisfy at least one covering pair, meaning that there is a
covering pair of $P$, $(i,j)$, such that $\perm(i) > \perm(j)$, so
that
\begin{equation*}
\prod_{i',j' : \perm(i') < \perm(j')} \e^{\quant^k_{i',j'} / b} \leq (n!)^{k-1} ~.
\end{equation*}
Since the total number of permutations is $n!$, the total sum of
\emph{all} terms in the sum $D_k$ corresponding to permutations that
do not satisfy the partial ordering $P$ is less than or equal to $n!
(n!)^{k-1} = (n!)^k$, and is strictly less than $(n!)^k$ unless the
number of linear extensions is 0, while the total sum of all the terms
corresponding to permutations that do satisfy $P$ is $\linexts(P)
(n!)^k$. Thus $\linexts(P) = \floor{D_k / (n!)^k}$.

We know that computing the number of linear extensions of a partial
ordering is $\SP$-hard. Therefore, both computing the prices and computing the value of the cost function in pair betting are $\SP$-hard.
\end{proof}

\begin{corollary}
Computing the payment of a transaction in a LMSR for pair betting is $\SP$-hard.
\end{corollary}
The proof is nearly identical to the proof of Corollary~\ref{cor:subset}.

\ignore{
\begin{proof}
If the market maker can compute $C(\tilde{Q}) - C(Q)$ for any quantities $\tilde{Q}$ and $Q$, it can compute $C(Q) - C(Q^0)$ for any $Q$. When the market maker starts the market with 0 share of every security, $C(Q^0) = b\log n!$ and the market maker is able to compute $C(Q)$ for any $Q$. According to Theorem \ref{T:pairprice}, computing the payment of a transaction in pair betting is $\SP$-hard.
\end{proof}
}

\section{LMSR for Boolean Betting}
We now examine an alternate type of market combinatorics in which the final outcome is a conjunction of event outcomes. Formally, let $\events$ be event space, consisting of $N$
individual events $\event_1,\cdots,\event_N$, which may or may not
be mutually independent. We define the state space $\states$ be the set of all possible joint outcomes for the $N$ events, so that its size is $|\states| = 2^N$. A Boolean betting market allows traders to bet on Boolean formulas of these events and their negations. A security $\angles{\phi}$ pays off \$1 if the Boolean formula $\phi$ is satisfied by the final outcome and \$0 otherwise.
For example, a security
$\angles{A_1 \lor A_2}$ pays off $\$1$ if and only if at least one
of events $A_1$ and $A_2$ occurs, while a security $\angles{A_1 \land A_3 \land
\lnot A_5}$ pays off $\$1$ if and only if the events $A_1$ and $A_3$
both occur and the event $A_5$ does not. Following the notational
conventions of Fortnow et al.~\cite{Fortnow04}, we use $\state \in
\phi$ to mean that the outcome $\state$ satisfies the Boolean formula $\phi$.
Similarly, $\state \not\in \phi$ implies that the outcome $\state$
does not satisfy $\phi$.

In this section, we focus our attention to LMSR markets for a very simple Boolean betting language, Boolean formulas of two events. We show that even when bets are only allowed to be placed on disjunctions or conjunctions of two events, it is still $\SP$-hard to calculate the prices, the value of the cost function, and payments of transactions in a Boolean betting market operated by a LMSR market maker.

Let $\setX$ be the set containing all elements of $\events$ and
their negations.  In other words, each event outcome $X_i \in
\setX$ is either $\event_j$ or $\lnot \event_j$ for some $\event_j
\in \events$. We begin by considering the scenario in which traders
may only trade securities $\angles{X_i \lor X_j}$ corresponding to disjunctions of any two event outcomes.

Let $q_{i,j}$ be the total number of shares purchased by all traders for the security
$\angles{X_i \lor X_j}$, which pays off $\$1$ in the event of any
outcome $\state$ such that $\state \in (X_i \lor X_j)$ and $\$0$
otherwise.  From Equation~\ref{eq:lmsr}, we can calculate the instantaneous price for the security $\angles{X_i \lor X_j}$ for any two event
outcomes $X_i, X_j \in \setX$ as
\begin{eqnarray}
\lefteqn{\price_{i,j}(Q)=}
\nonumber
\\
&\frac{\sum_{\state \in \states : \state \in (X_i \lor X_j)}
\prod_{1 \leq i' < j' \leq 2N : \state \in (X_{i'} \lor X_{j'})}
\e^{\quant_{i',j'} / b}}
{\sum_{\tau \in \states} \prod_{1 \leq i'
< j' \leq 2N : \tau \in (X_{i'} \lor X_{j'})} \e^{\quant_{i',j'} /
b}} .
\label{eqn:booleanprice}
\end{eqnarray}
Note that if $X_i = \lnot X_j$, $p_{i,j}(Q)$ is always $\$1$ regardless
of how many shares of other securities have been purchased. According to Equation~\ref{E:lmsr-cost2}, the cost function is
\begin{equation}
C(Q) = b \log {\sum_{\state \in \states} \prod_{1 \leq i
< j \leq 2N : \state \in (X_{i} \lor X_{j})} \e^{\quant_{i,j} /
b}}~.
\label{eqn:booleancost}
\end{equation}
Theorem \ref{thm:booleanhard} shows that computing prices and the value of the cost function in such
a market is $\SP$-hard, via a reduction from the \#2-SAT problem.\footnote{This can also be proved via a reduction from counting linear extensions using a similar technique to the proof of Theorem~\ref{thm:pairhard}, but the reduction to \#2-SAT is more natural.}

\begin{theorem}
It is $\SP$-hard to compute instantaneous prices in a LMSR market for Boolean betting when bets are restricted to disjunctions of two event outcomes.  Additionally, it is $\SP$-hard to compute the value of the cost function in this setting.
\label{thm:booleanhard}
\end{theorem}
\begin{proof}
Suppose we are given a 2-CNF (Conjunctive Normal Form) formula
\begin{equation}\label{eq:2cnf}
(X_{i_1} \lor X_{j_1}) \land (X_{i_2} \lor X_{j_2}) \land \cdots
\land (X_{i_k} \lor X_{j_k})
\end{equation}
with $k$ clauses, where each clause is a disjunction of two literals
(i.e. events and their negations).  Assume any redundant terms have
been removed.

The structure of the proof is similar to that of the pair betting case. We consider a Boolean betting markets with $N$ events, and show how to construct a sequence of trades that provides, through prices or the value of the cost function, the number of satisfiable assignments for the 2-CNF formula.

We create $k$ trading periods. At period $t$, a quantity $b \ln(2^N)$ of the security $\angles{X_{i_t} \lor X_{j_t}}$ is purchased. We denote by $p^t_{i,j}$ and $q^t_{i,j}$ respectively the price and outstanding quantities of the security $\angles{X_{i} \lor X_{j}}$ at the end of period $t$. Suppose the market starts with 0 share of every security. Note that $q^s_{i_t, j_t} = 0$ if $s < t$ and $q^s_{i_t, j_t} = b \ln(2^N)$ if $s \geq t$.
Let
\[
N_t(i,j) = {\sum_{\state \in \states : \state \in (X_i \lor X_j)} \prod_{1 \leq i' < j' \leq 2N : \state \in (X_{i'} \lor X_{j'})} \e^{\quant^t_{i',j'} / b}}~,
\]
and
\[
D_t = {\sum_{\state \in \states} \prod_{1 \leq i'
< j' \leq 2N : \state \in (X_{i'} \lor X_{j'})} \e^{\quant^t_{i',j'} /
b}}~.
\]
Thus, $p^t_{i,j} = N_t(i_t,j_t)/D_t$.

Since only one security $\angles{X_{i_1} \lor X_{j_1}}$ has been purchased in period 1, we get
\begin{eqnarray*}
\denom_1 &=& \sum_{\state \in \states : \state \in (X_{i_1} \lor
X_{j_1})} 2^N + \sum_{\state \in \states : \state \not\in (X_{i_1}
\lor X_{j_1})} 1
\\
&=& 3\cdot 2^{2N-2} + 2^{N-2}.
\end{eqnarray*}

We then show that $\denom_k$ can be calculated inductively from $\denom_1$. As the only security purchased in period $t$ is $(X_{i_t} \lor X_{j_t})$ in quantity $b \ln(2^N)$, we obtain
\[
\num_t(i_t, j_t) = \num_{t-1}(i_{t}, j_{t}) \e^{b \ln(2^N) / b} =
\num_{t-1}(i_{t}, j_{t}) 2^N~.
\]
Therefore,
\[
\frac{p^{t}_{i_t, j_t}}{p^{t-1}_{i_t, j_t}} =\frac{\num_t(i_t, j_t)
/ D_t}{\num_{t-1}(i_{t}, j_{t}) / D_{t-1}} =\frac{2^N D_{t-1}}{D_t},
\]
and we get
\begin{equation*}
D_k = (2^N)^{k-1} \left(\prod_{\ell=2}^{k} \frac{p^{\ell - 1}_{i_\ell, j_\ell}}{p^\ell_{i_\ell, j_\ell}}\right)D_1 ~.
\end{equation*}
In addition, since the cost function at the end of period $k$ can be expressed as
\[
C(Q) = b\log D_k~,
\]
$D_k$ can also be computed efficiently from the cost function in period $k$.

We now show that we can deduce from $D_k$ the number of satisfiable assignments for the 2-CNF formula (Equation~\ref{eq:2cnf}). Indeed, each term in the sum
\[
{\sum_{\state \in \states} \prod_{1 \leq i'
< j' \leq 2N : \state \in (X_{i'} \lor X_{j'})} \e^{\quant^k_{i',j'} /
b}}
\]
that corresponds to an outcome $\omega$ that satisfies the formula is exactly $2^{kN}$, as exactly $k$ terms in the product are $2^N$ and the rest is 1. On the contrary, each term in the sum that corresponds to an outcome $\state$ that does \emph{not} satisfy the 2-CNF formula will be at most $2^{(k-1)N}$ since at most $k-1$ terms in the product will be $2^N$ and the rest will be 1.  Since the total number of outcomes is $2^N$, the total sum of \emph{all} terms corresponding to outcomes that do not satisfy (\ref{eq:2cnf}) is less than or equal to $2^N
(2^{(k-1)N}) = 2^{kN}$, and is strictly less than $2^{kN}$ unless the number of satisfying assignments is 0.  Thus the number of
satisfying assignments is $\floor{D_k / 2^{kN}}$.

We know that computing the number of satisfiable assignments of a 2-CNF formula is \#P-hard. We have shown how to compute it in polynomial time using prices or the value of the cost function in a Boolean betting market of $N$ events. Therefore, both computing prices and computing the value of the cost function in a Boolean betting market is \#P-hard.
\end{proof}
\begin{corollary}\label{C:boolean}
Computing the payment of a transaction in a LMSR for Boolean betting is $\SP$-hard when traders can only bet on disjunctions of two events.
\end{corollary}
The proof is nearly identical to the proof of Corollary~\ref{cor:subset}.

\ignore{
\begin{proof}
If the market maker can compute $C(\tilde{Q}) - C(Q)$ for any quantities $\tilde{Q}$ and $Q$, it can compute $C(Q) - C(Q^0)$ for any $Q$. When the market maker starts the market with 0 share of every security, $C(Q^0) = b\log 2^N$ and the market maker is able to compute $C(Q)$ for any $Q$. According to Theorem \ref{thm:booleanhard}, computing the payment of a transaction in Boolean betting is $\SP$-hard.
\end{proof}
}

If we impose that participants in a Boolean betting market may only trade securities corresponding to conjunctions of any two event outcomes, $\angles{A_i \land A_j}$, the following Corollary gives the complexity results for this situation.

\begin{corollary}
It is $\SP$-hard to compute instantaneous prices in a LMSR market for Boolean betting when bets are restricted to conjunctions of two event outcomes.  Additionally, it is $\SP$-hard to compute the value of the cost function in this setting, and $\SP$-hard to compute the payment for a transaction.
\end{corollary}
\begin{proof}
Buying $q$ shares of security $\angles{A_i \land A_j}$ is equivalent to selling $q$ shares of $\angles{\lnot A_i \lor \lnot A_j}$.  Thus if we can operate a Boolean betting market for securities of the type $\angles{A_i \land A_j}$ in polynomial time, we can also operate a Boolean betting market for securities of the type $\angles{A_i \lor A_j}$ in polynomial time.  The result then follows from Theorem~\ref{thm:booleanhard} and Corollary~\ref{C:boolean}.
\end{proof}

\ignore{

\section{LMSR for Boolean Betting (Version 2: Linear Extensions Reduction)}

We consider Boolean betting of $N$ events where the only allowable
type of bets is conjunction of two events, e.g. $A_i \cap A_j$.  The
following theorem shows that pricing in LMSR for such Boolean
betting is $\SP$-hard.

\begin{theorem}
Pricing Boolean betting with LMSR when bets are restricted to
conjunctions of two events is $\SP$-hard.
\end{theorem}
\begin{proof}
The proof is by reduction from the problem of computing the number
of linear extensions to any partial ordering.

First, we transform any partial ordering into a sequence of Boolean
betting markets as follows.
\begin{enumerate}
\item
Let $P$ represents a partial ordering over $n$ elements with $k$
relations. $N(P)$ is the number of linear extensions of $P$. Let
$P'$ represents the partial ordering over $n$ elements with $k'$
relations.  $P'$ is obtained by adding those relations implied by
transitivity to $P$. For example, if $P$ contains two relations
$a<b$ and $b<c$, $P'$ contains three relations $a<b$, $b<c$, and
$a<c$. Thus, $N(P') = N(P')$.
\item
Let $(i_1, j_1), (i_2, j_2), ... , (i_{k'}, j_{k'})$ be the $k'$
relations in $P'$. $(i_m, j_m)$ means that  $i_m < j_m$. We
construct a sequence of partial orderings, $P^1$, $P^2$, ... ,
$P^{k'-1}$, over the $n$ elements, with $P^m$ contains the first
$m+1$ relations in $P'$ for $m \in \{1, ..., (k' -1)\}$. Hence,
$P^{k'-1} = P'$.
\item
For the $n$ elements, there are a total number of $n(n-1)/2$
relational pairs. Let $N=n(n-1)$. Each relational pair corresponds
to a binary event. For example, the event for $(a, b)$ has outcome
$a<b$ or $a\geq b$. We create a sequence of Boolean betting markets,
$M^1$, $M^2$, ... , $M^{k'-1}$, for these $N$ events, corresponding
to $P^1$, $P^2$, ... , $P^{k'-1}$. In market $M^m$, traders buy $q=
b \log (2^N)$ shares of $(i_s< j_s) \cap (i_{s+1} < j_{s+1}) $ for
every $s \in \{1, 2, ... , m\}$ and nothing else.
\end{enumerate}

Second, we show that we can calculate the denominators of the price
function in the sequence of markets that we setup, if we can price
Boolean betting. The price for security $(i_s< j_s) \cap (i_{s+1} <
j_{s+1}) $ in a Boolean betting market is
\begin{equation}
p_{s, s+1} =\frac{ \sum_{\omega \in \{(i_s< j_s) \cap (i_{s+1} <
j_{s+1})\}} \e^{q(\omega)/b}}{\sum_{\omega \in \Omega}
\e^{q(\omega)/b}}
\end{equation}
Let $p_{s, s+1}^m$ denote the price for security $(i_s< j_s) \cap
(i_{s+1} < j_{s+1})$ in market $m$. Let  $D^m$ represent the
denominator of the price function in market $m$. Let $N_{s, s+1}^m$
be the numerator of the price function for security $(i_s< j_s) \cap
(i_{s+1} < j_{s+1})$ in market $m$. We can see that $p_{s,
s+1}^m=N_{s, s+1}^m/D^m$. For market $M^1$, we can calculate $D^1 =
2^{N-2}\e^{q/b}+(2^N -2^{N-2})\e^{0/b}=2^{2N-2} + 2^N -2^{N-2}$.
Consider the price of security $(i_s< j_s) \cap (i_{s+1} < j_{s+1})$
in market $M^s$ and $M^{s-1}$ for $s>1$,
\[
\frac{p^s_{s, s+1}}{p^{s-1}_{s, s+1}}= \frac{\frac{N^s_{s,
s+1}}{D^s}}{\frac{N^{s-1}_{s,
s+1}}{D^{s-1}}}=\frac{D^{s-1}}{D^s}\frac{N^s_{s, s+1}}{N^{s-1}_{s,
s+1}}=\frac{D^{s-1}}{D^s}\frac{N^{s-1}_{s, s+1}\e^{q/b}}{N^{s-1}_{s,
s+1}}=\frac{2^N D^{s-1}}{D^s}.
\]
From the above equation, we can solve
\[
D^s = \frac{2^N D^{s-1} p^{s-1}_{s, s+1}}{p^s_{s, s+1}}.
\]
By induction, we will be able to calculate $D^{k'-1}$ is we can
price Boolean betting.

Finally, we show that if we can calculate $D^{k'-1}$, we can
calculate $N(P)$. For each linear extension of $P$, there is a
corresponding state $\omega$ such that it's quantity $q(\omega) =
(k'-1)q$, meaning that $\e^{q(\omega)/b} = 2^{N(k'-1)}$. If a state
$\omega'$ is not a linear extension of $P$, $q(\omega')$ is at most
$(k'-2)q$ and $\e^{q(\omega')/b}$ is at most $2^{N(k'-2)}$. There are
at most $2^N$ such $\omega'$ that are not a linear extensions of
$P$, hence the sum of $\e^{q(\omega')/b}$ is less than $2^{N(k'-1)}$.
By dividing $D^{k'-1}$ by $2^{N(k'-1)}$ and drop the remainder, we
get $N(P)$.

Since calculating $N(P)$ is $\SP$-hard, pricing Boolean betting when
only conjunctions of two events are allowed is $\SP$-hard.
\end{proof}

\begin{theorem}
Pricing Boolean betting when only conjunctions of $k$ events are
allowed ($k>1$) is $\SP$-hard.
\end{theorem}
\begin{proof}
This comes easily by induction. Because if we can price Boolean
betting when conjunction of $k$ events are allowed, we can price
Boolean betting of $k-1$ events.
\end{proof}

} 

\section{An Approximation Algorithm for Subset Betting}

There is an interesting relationship between logarithmic market
scoring rule market makers and a common class of algorithms for
online learning in an experts setting.  In this section, we
elaborate on this connection, and show how results from the online
learning community can be used to prove new results about an
approximation algorithm for subset betting.

\subsection{The Experts Setting}

We begin by describing the standard model of online learning with
expert advice~\cite{Littlestone94, Freund97, Vovk95}.  In this model,
at each time $t \in \{1,\cdots,T\}$, each expert $i \in \{1,\cdots,
n\}$ receives a \emph{loss} $\loss_{i,t}\in [0,1]$.  The
\emph{cumulative loss} of expert $i$ at time $T$ is $\Loss_{i,T} =
\sum_{t=1}^T \loss_{i,t}$. No statistical assumptions are made about these losses, and in general, algorithms are expected to perform well even if the sequence of losses is chosen by an adversary.

An algorithm $\alg$ maintains a current weight $w_{i,t}$ for each expert $i$, where $\sum_{i=1}^n w_{i,t} = 1$. These
weights can be viewed as distributions over the experts.  The
algorithm then receives its own instantaneous loss $\loss_{\alg,t}
=\sum_{i=1}^n w_{i,t} \loss_{i,t}$, which may be interpreted as the expected loss of the algorithm when choosing an expert according to the current distribution. The cumulative loss of $\alg$ up to time $T$ is then defined in the natural way as
$\Loss_{\alg,T} = \sum_{t=1}^T \loss_{\alg,t}=\sum_{t=1}^T
\sum_{i=1}^n w_{i,t} \loss_{i,t}$.  A common goal in such online learning settings is to minimize an algorithm's \emph{regret}. Here
the regret is defined as the difference between the cumulative loss of the
algorithm and the cumulative loss of an algorithm that would have ``chosen'' the best expert in hindsight by setting his weight to $1$ throughout all the periods. Formally, the regret is given by $\Loss_{\alg,T} - \min_{i \in \{1,\cdots,n\}} \Loss_{i,T}$.


Many algorithms that have been analyzed in the online experts setting
are based on exponential weight updates.  These exponential updates
allow the algorithm to quickly transfer weight to an expert that is
outperforming the others. For example, in the Weighted Majority
algorithm of Littlestone and Warmuth~\cite{Littlestone94}, the weight
on each expert $i$ is defined as
\begin{eqnarray}\label{eqn:wmupdate}
w_{i,t}
= \frac{w_{i,t-1} \e^{-\eta \loss_{i,t}}}
{\sum_{j=1}^n w_{j,t-1} \e^{-\eta \loss_{j,t}}}
= \frac{\e^{-\eta \Loss_{i,t}}} {\sum_{j=1}^n \e^{-\eta
\Loss_{j,t}}} ~,
\end{eqnarray}
where $\eta$ is the \emph{learning rate}, a small positive parameter
that controls the magnitude of the updates.  The following theorem
gives a bound on the regret of Weighted Majority.  For a proof of this result and a nice overview of learning with expert advice, see, for example, Cesa-Bianchi and Lugosi~\cite{Cesa06}.

\begin{theorem}\label{thm:wm}
Let $\alg$ be the Weighted Majority algorithm with parameter $\eta$. After a sequence of $T$ trials,
\[
\Loss_{\alg,T} - \min_{i \in \{1,\cdots,n\}} \Loss_{i,T}
\leq
\eta T + \frac{\ln(n)}{\eta} ~.
\]
\end{theorem}

\subsection{Relationship to LMSR Markets}
\label{sec:expertrelation}

There is a manifest similarity between the expert weights used by Weighted Majority and the prices in the LMSR market.  One might ask if the results from the experts setting can be applied to the analysis of prediction markets. Our answer is \emph{yes}. In fact, it is possible to use Theorem~\ref{thm:wm} to rediscover the well-known bound of $b \ln(n)$ for the loss of an LMSR market maker with $n$ outcomes.

Let $\epsilon$ be a limit on the number of shares that a trader may purchase or sell at each time step; in other words, if a trader would like to purchase or sell $q$ shares, this purchase must be broken down into $\ceil{q/\epsilon}$ separate purchases of $\epsilon$ or less shares.  Note that the total number of time steps $T$ needed to execute such a sequence of purchases and sales is proportional to $1/\epsilon$.

We will construct a sequence of loss functions in a setting with $n$ experts to induce a sequence of weight matrices that correspond to the price matrices of the LMSR market. At each time step $t$, let $p_{i,t} \in [0,1]$ be the instantaneous price of security $i$ at the end of period $t$, and let $q_{i,t} \in [-\epsilon,\epsilon]$ be the number of shares of security $i$ purchased during period $t$.  Let $Q_{i,t}$ be the total number of shares of security $i$ that have been purchased up to time $t$.  Now, let's define the instantaneous loss of each expert as
$\loss_{i,t} = (2\epsilon - q_{i,t})/(\eta b)$.
First notice that this loss is always in $[0,1]$ as long as $\eta \geq 2\epsilon/b$.  From Equations~\ref{E:lmsr-price} and~\ref{eqn:wmupdate}, at each time $t$,
\begin{eqnarray*}
p_{i,t}
&=& \frac{ \e^{Q_{i,t}/b}}
{\sum_{j=1}^n \e^{Q_{j,t}/b}}
= \frac{ \e^{2\epsilon t/b - \eta \Loss_{i,t}}}
{\sum_{j=1}^n \e^{2\epsilon t/b - \eta \Loss_{j,t}}}
\\
&=& \frac{\e^{- \eta \Loss_{i,t}}}
{\sum_{j=1}^n \e^{- \eta \Loss_{j,t}}} = w_{i,t} ~.
\end{eqnarray*}

Applying Theorem~\ref{thm:wm}, and rearranging terms, we find that
\[
\max_{i \in \{1,\cdots,n\}} \sum_{t=1}^T q_{i,t} -
\sum_{t=1}^T \sum_{i=1}^n p_{i,t} q_{i,t}
\leq \eta^2 T b + b \ln(n).
\]
The first term of the left-hand side is the maximum payment that the market maker needs to make, while the second terms of the left-hand side captures the total money the market maker has received.
The right hand side is clearly minimized when $\eta$ is set as small as possible.  Setting $\eta = 2\epsilon/b$ gives us
\[
\max_{i \in \{1,\cdots,n\}} \sum_{t=1}^T q_{i,t} -
\sum_{t=1}^T \sum_{i=1}^n p_{i,t} q_{i,t}
\leq 4 \epsilon^2 T b + b \ln(n).
\]

Since $T = O(1/\epsilon)$, the term $4\epsilon^2 T b$ goes to 0 as $\epsilon$ becomes very small.  Thus in the limit as $\epsilon \rightarrow 0$, we get the well-known result that the worst-case loss of the market maker is bounded by $b \ln(n)$.

\subsection{Considering Permutations}

Recently Helmbold and Warmuth~\cite{Helmbold07} have shown that many
results from the standard experts setting can be extended to a setting
in which, instead of competing with the best expert, the goal is to compete with the best permutation over $n$ items.  Here each permutation suffers a loss at each time step, and the goal of the algorithm is to maintain a weighting over permutations such that the cumulative regret
to the best permutation is small.  It is infeasible to treat each
permutation as an expert and run a standard algorithm since this would
require updating $n!$ weights at each time step.  Instead, they show
that when the loss has a certain structure (in particular, when the
loss of a permutation is the sum of the losses of each of the $n$
mappings), an alternate algorithm can be used that requires tracking
only $n^2$ weights in the form of an $n \times n$ doubly stochastic
matrix.

Formally, let $W^t$ be a doubly stochastic matrix of weights
maintained by the algorithm $\alg$ at time $t$.  Here $W^t_{i,j}$ is
the weight corresponding to the probability associated with item $i$ being mapped into
position $j$.  Let $L^t \in [0,1]^{n\times n}$ be the loss matrix at
time $t$.  The instantaneous loss of a permutation $\perm$ at time
$t$ is $\loss_{\perm,t} = \sum_{i=1}^n L^t_{i,\perm(i)}$.  The
instantaneous loss of $\alg$ is $\loss_{\alg,t} = \sum_{i=1}^n
\sum_{j=1}^n W^t_{i,j} L^t_{i,j}$, the matrix dot product between
$W^t$ and $L^t$.  Notice that $\loss_{\alg,t}$ is equivalent to the
expectation over permutations $\perm$ drawn according to $W^t$ of
$\loss_{\perm,t}$. The goal of the algorithm is to minimize the cumulative regret to the best permutation, $\Loss_{\alg,T} - \min_{\perm \in \Omega} \Loss_{\perm,T}$ where the cumulative loss is defined as before.

Helmbold and Warmuth go on to present an algorithm called PermELearn
that updates the weight matrix in two steps.  First, it creates a
temporary matrix $W'$, such that for every $i$ and $j$, $W'_{i,j} =
W^{t}_{i,j} \e^{-\eta L^t_{i,j}}$.  It then obtains $W^{t+1}_{i,j}$ by
repeatedly rescaling the rows and columns of $W'$ until the matrix is
doubly stochastic.
Alternately rescaling rows and columns of a matrix $M$ in this way is known as Sinkhorn balancing~\cite{Sinkhorn64}. Normalizing the rows of a matrix is equivalent to
pre-multiplying by a diagonal matrix, while normalizing the columns is
equivalent to post-multiplying by a diagonal matrix.  Sinkhorn~\cite{Sinkhorn64} shows that this procedure converges to a unique
doubly stochastic matrix of the form $RMC$ where $R$ and $C$ are
diagonal matrices if $M$ is a positive matrix.  Although there are cases in which Sinkhorn
balancing does not converge in finite time, many results show that the number of Sinkhorn iterations needed to scale a matrix so that row and column sums are $1\pm \epsilon$ is polynomial in $1/\epsilon$~\cite{Hala04, Kalantari96, Linial00}.

The following theorem~\cite{Helmbold07} bounds the cumulative loss of
the PermELearn in terms of the cumulative loss of the best
permutation.

\begin{theorem}{\sl(Helmbold and Warmuth~\cite{Helmbold07})}
Let $\alg$ be the PermELearn algorithm with parameter $\eta$.  After
a sequence of $T$ trials,
\[
\Loss_{\alg,T} \leq \frac{n \ln(n) + \eta \min_{\perm \in \Omega}
\Loss_{\perm,T}} {1 - \e^{-\eta}} ~.
\]
\label{thm:permelearn}
\end{theorem}

\subsection{Approximating Subset Betting}

Using the PermELearn algorithm, it is simple to approximate prices for
subset betting in polynomial time.  We start with a $n \times n$ price
matrix $P^1$ in which all entries are $1/n$.  As before, traders may
purchase securities of the form $\angles{i|\Phi}$ that pay off $\$1$
if and only if horse or candidate $i$ finishes in a position $j \in
\Phi$, or securities of the form $\angles{\Psi|j}$ that pay off $\$1$
if and only if a horse or candidate $i \in \Psi$ finishes in position
$j$.

As in Section~\ref{sec:expertrelation}, each time a trader purchases or sells $q$ shares, the purchase or sale is broken up into $\ceil{q/\epsilon}$ purchases or sales of $\epsilon$ shares or less, where $\epsilon > 0$ is a small constant.\footnote{We remark that dividing purchases in this way has the negative effect of creating a polynomial time dependence on the quantity of shares purchased. However, this is not a problem if the quantity of shares bought or sold in each trade is bounded to start, which is a reasonable assumption.  The additional time required is then linear only in $1/\epsilon$.}
Thus we can treat the sequence of purchases as a sequence of $T$ purchases of $\epsilon$ or less shares, where $T = O(1/\epsilon)$.
Let $q^t_{i,j}$ be the number of shares of securities
$\angles{i|\Phi}$ with $j \in \Phi$ or $\angles{\Psi|j}$ with $i \in
\Psi$ purchased at time $t$; then $q^t_{i,j} \in [-\epsilon,\epsilon]$
for all $i$ and $j$.

The price matrix is updated in two steps.  First, a temporary matrix
$P'$ is created where for every $i$ and $j$, $P'_{i,j} = P^t_{i,j}
\e^{q^t_{i,j}/b}$ where $b > 0$ is a parameter playing a similar role to $b$ in Equation~\ref{E:lmsr-price}.  Next, $P'$ is Sinkhorn balanced to the desired precision, yielding an (approximately) doubly stochastic matrix
$P^{t+1}$.

The following lemma shows that updating the price matrix in this way
results in a price matrix that is equivalent to the weight matrix of
PermELearn with particular loss functions.

\begin{lemma}
The sequence of price matrices obtained by the approximation algorithm
for subset betting on a sequence of purchases $q^t \in
[-\epsilon,\epsilon]^{n \times n}$ is equivalent to the sequence of
weight matrices obtained by running PermELearn($\eta$) on a sequence
of losses $L^t$ with
\[
L^t_{i,j} =\frac{2\epsilon - q^t_{i,j}}{\eta b}
\]
for all $i$ and $j$, for any $\eta \geq 2\epsilon/b$.
\label{lemma:permelearnsubset}
\end{lemma}
\begin{proof}
First note that for any $\eta \geq 2\epsilon/b$, $L^t_{i,j} \in [0,1]$ for all $t$, $i$, and $j$, so the loss matrix is valid for
PermELearn.  $P^1$ and $W^1$ both contain all entries of $1/n$.
Assume that $P^t = W^t$. When updating weights for time $t+1$, for all
$i$ and $j$,
\begin{eqnarray*}
P'_{i,j} &=&  P^t_{i,j} \e^{q^t_{i,j}/b} = W^{t}_{i,j} \e^{q^t_{i,j}/b}
= \e^{2\epsilon / b} W^{t}_{i,j} \e^{-2\epsilon/b + q^t_{i,j}/b }
\\
&=& \e^{2\epsilon/b} W^{t}_{i,j} \e^{-\eta L^t_{i,j}}
= \e^{2\epsilon/b} W'_{i,j} ~.
\end{eqnarray*}
Since the matrix $W'$ is a constant multiple of $P'$, the
Sinkhorn balancing step will produce the same matrices.
\end{proof}

Using this lemma, we can show that the difference between the amount
of money that the market maker must distribute to traders in the worst
case (i.e. when the true outcome is the outcome that pays off the
most) and the amount of money collected by the market is bounded.  We
will see in the corollary below that as $\epsilon$ approaches 0, the
worst case loss of the market maker approaches $bn \ln(n)$, regardless
of the number of shares purchased.  Unfortunately, if $\epsilon > 0$,
this bound can grow arbitrarily large. 


\begin{theorem}\label{thm:approxbound}
For any sequence of valid subset betting purchases $q^{t}$ where
$q^t_{i,j} \in [-\epsilon,\epsilon]$ for all $t$, $i$, and $j$, let
$P^1, \cdots, P^T$ be the price matrices obtained by running the
subset betting approximation algorithm.  Then
\begin{eqnarray*}
\lefteqn{
\max_{\perm \in \perms}
\sum_{t=1}^T \sum_{i=1}^n  q^t_{i,\perm(i)}
- \sum_{t=1}^T \sum_{i=1}^n \sum_{j=1}^n P^t_{i,j} q^t_{i,j}
}
\\
&\leq& \frac{2\epsilon/b}{1 - \e^{-2\epsilon/b}} n \ln(n)
+ \left(\frac{2\epsilon/b}{1 - \e^{-2\epsilon/b}} - 1\right) 2\epsilon nT ~.
\end{eqnarray*}
\end{theorem}
\begin{proof}
By Theorem~\ref{thm:permelearn} and Lemma~\ref{lemma:permelearnsubset}, after rearranging terms, we have that for any $\eta \geq 2\epsilon/b$,
\begin{eqnarray*}
\lefteqn{
2\epsilon nT - \sum_{t=1}^T \sum_{i=1}^n \sum_{j=1}^n P^t_{i,j}
}
\\
&\leq&
\left(\frac{\eta}{1-\e^{-\eta}}\right)
\left(b n\ln n  + 2\epsilon nT -
\max_{\perm \in \perms}  \sum_{t=1}^T \sum_{i=1}^n q^t_{i,\perm(i)}  \right) ~.
\end{eqnarray*}

Thus we have
\begin{eqnarray*}
\lefteqn{
\frac{\eta}{1 - \e^{-\eta}} \max_{\perm \in \perms}
\sum_{t=1}^T \sum_{i=1}^n  q^t_{i,\perm(i)}
- \sum_{t=1}^T \sum_{i=1}^n \sum_{j=1}^n P^t_{i,j} q^t_{i,j}
}
\\
&\leq& \frac{\eta}{1 - \e^{-\eta}} b n \ln(n)
+ \left(\frac{\eta}{1 - \e^{-\eta}} - 1\right) 2\epsilon nT ~.
\end{eqnarray*}

Notice that $\eta/(1-\e^{-\eta})$ is an increasing function in $\eta$,
and goes to $1$ in the limit as $\eta$ goes to 0.  Thus the right hand
side of this equation decreases as $\eta$ decreases to 0.  Setting
$\eta = 2\epsilon/b$ yields the result.
\end{proof}

Notice that the number of steps $T$ scales inversely with
$\epsilon$ since each lump purchase of $q$ shares must be broken into
$\ceil{q/\epsilon}$ individual purchases.  Thus in the limit as
$\epsilon \rightarrow 0$, the loss of the market maker is bounded by
$b n\ln(n)$.

\begin{corollary}
For any sequence of valid subset betting purchases broken into $T$
($= O(1/\epsilon)$) small purchases such that $q^t_{i,j} \in
[-\epsilon,\epsilon]$ for all $t$, $i$, and $j$, let $P^1, \cdots,
P^T$ be the price matrices obtained by running the Subset Betting
Approximation Algorithm.  Then in the limit as $\epsilon \rightarrow
0$,
\[
\max_{\perm \in \perms} \sum_{t=1}^T \sum_{i=1}^n  q^t_{i,\perm(i)}
- \sum_{t=1}^T \sum_{i=1}^n \sum_{j=1}^n P^t_{i,j} q^t_{i,j}
\leq b n \ln(n) ~.
\]
\end{corollary}

This bound is comparable to worst-case loss bounds achieved using alternate methods for operating LMSRs on permutations.  A single LMSR operated on the entire outcome space has a guaranteed worst-case loss of $b\ln(n!)$, but is, of course, intractible to operate.  A set of $n$ LMSRs operated as $n$ separate markets, one for each position, would also have a total worst-case loss $bn\ln(n)$, but could not guarantee consistent prices. In the limit, our approximation algorithm achieves the same worst-case loss guarantee as if we were operating $n$ separate markets, but prices remain consistent at all times.

\section{Conclusions}

We investigate the computational complexity for logarithmic market scoring rule (LMSR) market makers to operate combinatorial betting markets. We examine two specific market combinatorics, permutations and Boolean combinations.  In a permutation betting market, the state space is the set of rankings of $n$ competing candidates and is of size of $n!$.  In a Boolean betting market, the state space is the set of Boolean combinations of $N$ event outcomes and is of size $2^N$.

Since allowing participants to bet on individual states is both intractable and unnatural, we allow participants to trade securities that correspond to simple properties of the final state. For permutation betting, we consider subset betting, which allows traders to bet on a set of positions that a candidate will stand at or a set of traders who will stand at a particular position, and pair betting, which allows traders to wager on the relative ranking of two candidates. For Boolean betting, we consider the situation where traders bet on conjunctions or disjunctions of two events. In all cases, we prove that it is $\SP$-hard to compute the instantaneous prices as well as payments of transactions in a LMSR market.  Our results on subset betting in LMSR contrast with those of Chen et al.~\cite{Chen07} who show that subset betting is tractable when the market is operated by a central auctioneer who performs riskless order matching. Our results on Boolean betting contrast with those of Chen, Goel, and Pennock~\cite{Chen08} who consider betting on single-elimination tournaments---a special form of Boolean combinations---and show that such a market with LMSR may be operated in polynomial time. This raises interesting questions on the connection between the complexity of auctioneers and the complexity of market makers, and on the complexity of other combinatorial betting scenarios.

We also show that there is an interesting relationship between LMSR market makers and a common class of expert learning algorithms. This allows us to design an approximation algorithm for pricing subset betting in an LMSR. We prove that in the limit the worst-case loss of an LMSR market maker that uses our algorithm remains bounded. In the future we plan to further investigate the connection between online learning in expert settings and information markets with automated market makers.

\section{Acknowledgments}

The authors are grateful to Sampath Kannan for useful discussions and advice on how to approach the pair betting problem, to Manfred Warmuth for sharing extended details of his work on learning permutations, and to Sharad Goel for insightful discussions.

\bibliography{lmsr}

\begin{thebibliography}{10}

\bibitem{Hala04}
H.~Balakrishnan, I.~Hwang, and C.~Tomlin.
\newblock Polynomial approximation algorithms for belief matrix maintenance in
  identity management.
\newblock In {\em 43rd IEEE Conference on Decision and Control}, pages
  4874--4879, 2004.

\bibitem{Brightwell91}
G.~Brightwell and P.~Winkler.
\newblock Counting linear extensions is \#{P}-complete.
\newblock In {\em ACM Symposium on Theory of Computing}, 1991.

\bibitem{Cesa06}
N.~Cesa-Bianchi and G.~Lugosi.
\newblock {\em Prediction, learning, and games}.
\newblock Cambridge University Press, 2006.

\bibitem{Chen07}
Y.~Chen, L.~Fortnow, E.~V. Nikolova, and D.~M. Pennock.
\newblock Betting on permutations.
\newblock In {\em ACM Conference on Electronic Commerce}, 2007.

\bibitem{Chen07-3}
Y.~Chen, L.~Fortnow, E.~V. Nikolova, and D.~M. Pennock.
\newblock Combinatorial betting.
\newblock {\em ACM SIGecom Exchanges}, 7(1):865--877, 2007.

\bibitem{Chen08}
Y.~Chen, S.~Goel, and D.~M. Pennock.
\newblock Pricing combinatorial markets for tournaments.
\newblock In {\em ACM Symposium on Theory of Computing}, 2008.
\newblock To appear.

\bibitem{Chen07-2}
Y.~Chen, D.~M. Reeves, D.~M. Pennock, R.~D. Hanson, L.~Fortnow, and R.~Gonen.
\newblock Bluffing and strategic reticence in prediction markets.
\newblock In {\em Workshop on Internet and Network Economics}, 2007.

\bibitem{Chen07-4}
Yiling Chen and David~M. Pennock.
\newblock A utility framework for bounded-loss market makers.
\newblock In {\em Proceedings of the 23rd Conference on Uncertainty in
  Artificial Intelligence}, pages 49--56, 2007.

\bibitem{Forsythe92}
R.~Forsythe, F.~Nelson, G.~Neumann, and J.~Wright.
\newblock Anatomy of an experimental political stock market.
\newblock {\em The American Economic Review}, 82(5):1142--1161, 1992.

\bibitem{Forsythe99}
R.~Forsythe, T.~Rietz, and T.~Ross.
\newblock Wishes, expectations, and actions: {A} survey on price formation in
  election stock markets.
\newblock {\em Journal of Economic Behavior and Organization}, 39:83--110,
  1999.

\bibitem{Fortnow04}
L.~Fortnow, J.~Kilian, D.~M. Pennock, and M.~Wellman.
\newblock Betting boolean-style: {A} framework for trading securities based on
  logical formulas.
\newblock {\em Decision Support Systems}, 39(1):87--104, 2004.

\bibitem{Freund97}
Y.~Freund and R.~Schapire.
\newblock A decision-theoretic generalization of on-line learning and an
  application to boosting.
\newblock {\em Journal of Computer and System Sciences}, 55(1):119--139, 1997.

\bibitem{Hanson03}
R.~D. Hanson.
\newblock Combinatorial information market design.
\newblock {\em Information Systems Frontiers}, (1):105--119, 2003.

\bibitem{Hanson07}
R.~D. Hanson.
\newblock Logarithmic market scoring rules for modular combinatorial
  information aggregation.
\newblock {\em Journal of Prediction Markets}, 2007.

\bibitem{Hanson08}
R.~D. Hanson, J.~Ledyard, and T.~Ishikida.
\newblock An experimental test of combinatorial information markets.
\newblock {\em Journal of Economic Behavior and Organization}, 2008.
\newblock To appear.

\bibitem{Helmbold07}
D.~Helmbold and M.~Warmuth.
\newblock Learning permutations with exponential weights.
\newblock In {\em 20th Annual Conference on Learning Theory}, pages 469--483,
  2007.

\bibitem{Kalantari96}
Bahman Kalantari and Leonid Khachiyan.
\newblock On the complexity of nonnegative-matrix scaling.
\newblock {\em Linear Algebra and its applications}, (240):87--103, 1996.

\bibitem{Linial00}
Nathan Linial, Alex Samorodnitsky, and Avi Wigderson.
\newblock A deterministic strongly polynomial algorithm for matrix scaling and
  approximate permanents.
\newblock {\em Combinatorica}, 20(4):545--568, 2000.

\bibitem{Littlestone94}
N.~Littlestone and M.~Warmuth.
\newblock The weighted majority algorithm.
\newblock {\em Information and Computation}, 108(2):212--261, 1994.

\bibitem{Nelson89}
R.~Nelson and D.~Bessler.
\newblock Subjective probabilities and scoring rules: {E}xperimental evidence.
\newblock {\em American Journal of Agricultural Economics}, 71(2):363--369,
  1989.

\bibitem{Oliven04}
K.~Oliven and T.~Rietz.
\newblock Suckers are born, but markets are made: {I}ndividual rationality,
  arbitrage, and market efficiency on an electronic futures market.
\newblock {\em Management Science}, 50(3):336--351, 2004.

\bibitem{Sinkhorn64}
R.~Sinkhorn.
\newblock A relationship between arbitrary positive matrices and doubly
  stochastic matrices.
\newblock {\em The Annals of Mathematical Statistics}, 35(2):876--879, 1964.

\bibitem{Thaler88}
R.~Thaler and W.~Ziemba.
\newblock Anomalies: {P}arimutuel betting markets: {R}acetracks and lotteries.
\newblock {\em Journal of Economic Perspectives}, 2(2):161--174, 1988.

\bibitem{toda}
S.~Toda.
\newblock {PP} is as hard as the polynomial-time hierarchy.
\newblock {\em SIAM Journal on Computing}, 20(5):865--877, 1991.

\bibitem{Va}
L.~Valiant.
\newblock The complexity of computing the permanent.
\newblock {\em Theoretical Computer Science}, 8:189--201, 1979.

\bibitem{Valiant79}
L.~Valiant.
\newblock The complexity of enumeration and reliability problems.
\newblock {\em SIAM Journal of Computing}, 8(3):410--421, 1979.

\bibitem{Vovk95}
V.~Vovk.
\newblock A game of prediction with expert advice.
\newblock {\em Journal of Computer and System Sciences}, 56:153--173, 1998.

\end{thebibliography}
\bibliographystyle{plain}

\end{document}